\documentclass[sigconf]{acmart}

\AtBeginDocument{%
  }

\setcopyright{none}

\usepackage{booktabs}
\usepackage{multirow}
\usepackage{amsmath}
\usepackage{xcolor}

\begin{document}
\tolerance=1500
\emergencystretch=1em

\title{Do Recommendation Algorithms Work When Users Are LLM Agents? A Case Study on Moltbook}

\author{Daming Li}
\affiliation{%
  \institution{Independent Researcher}
  \city{Mountain View, CA}
  \country{USA}}
\email{damingliyale22@gmail.com}

\author{Simeng Han}
\affiliation{%
  \institution{Stanford University}
  \city{Palo Alto, CA}
  \country{USA}}
\email{shan6@stanford.edu}

\author{Jialu Zhang}
\authornote{Corresponding author.}
\affiliation{%
  \institution{University of Waterloo}
  \city{Waterloo, ON}
  \country{Canada}}
\email{jialu.zhang@uwaterloo.ca}

\begin{abstract}
Large language model (LLM) agents are increasingly populating web platforms, raising a fundamental question for recommender systems: do algorithms designed for human users still work when users are LLM agents that may not have well-defined content consumption preferences? We study this question by formulating a forum recommendation problem on Moltbook, a large-scale social media platform exclusively for autonomous AI agents running on the OpenClaw framework. We evaluate nine recommendation methods spanning simple heuristic rules, matrix factorization, item- and user-based collaborative filtering, graph-based, and sequential models on the task of predicting which forums an agent will engage with next.

We find that simple popularity-based rules or item-side collaborative filtering leveraging the platform and item structural information outperform techniques that explicitly learn a user representation. The static agent persona descriptions, the closest analog to a preference profile, fail to add value in predicting engagement. These results suggest that, on Moltbook, recommendation depends more on platform- and item-level structural signals than on user-specific personalization. We present multiple lines of empirical evidence that the observed content consumption patterns on Moltbook differ from well-established findings on human recommendation datasets, providing a new angle for studying agent societies and designing robust recommendation algorithms as agents increasingly populate the web.

\end{abstract}



\maketitle

\section{Introduction}
\label{sec:intro}

Recommender systems have been a cornerstone of modern web platforms for the past two decades, enabling users to discover relevant content by learning from their behavioral history~\cite{ricci2010introduction, bobadilla2013recommender, li2024recent}. From early collaborative filtering~\cite{goldberg1992using, hu2008collaborative, koren2009matrix, su2009survey} to modern deep learning~\cite{he2017neural, liang2018variational, zhang2019deep} and transformer-based sequential models~\cite{kang2018self, sun2019bert4rec, zhou2018deep, xia2023transact}, the field has made remarkable progress. The assumption underlying all these advances is that users possess latent yet learnable preferences that persist across interactions and evolve gradually over time. Human users' interests tend to be consistent within sessions and drift slowly across weeks or months~\cite{campos2014time, vinagre2015overview,bogina2023considering}, enabling increasingly sophisticated personalization.

However, a new class of web users, autonomous LLM agents~\cite{park2023generative, wang2024survey}, is emerging that may weaken or alter this assumption. These agents, powered by LLMs, are increasingly deployed on web platforms, where they perform tasks and consume content alongside human users~\cite{lombardi2022ai, durante2024agent, shin2026ai, zhou2026externalization}. Still, it is unclear whether they possess a clear content preference profile. Many deployed agent architectures~\cite{xi2025rise, yao2022react, shinn2023reflexion, abou2025agentic} have limited or no persistent memory across sessions---each conversation turn may reset the agent's context, and even agents configured with external memory tools (e.g., memory files, persona descriptions) may not develop a persistent yet evolving preference profile that recommender systems assume. Consider this scenario: when a human user encounters an interesting post from a category they have never explored before, this exposure can spark a new, lasting interest that shapes their future browsing and engagement patterns. An LLM agent, by contrast, is typically stateless between sessions, and it remains unclear whether agents exhibit persistent engagement patterns analogous to human preference development.

This raises a fundamental question for recommender systems and agent behavioral studies. If agent users lack clear preferences, can existing recommendation models---which fundamentally rely on identifying preference patterns---still be effective? Or, do we need entirely different approaches for agent-populated platforms? As LLM agents become more prevalent on the web, answering this question has both scientific and practical importance~\cite{ashery2025emergent, chen2026ai, zhou2026externalization}.

We investigate this question empirically on Moltbook\footnote{\url{https://www.moltbook.com/}}, the first recently launched large-scale social media platform exclusively designed for autonomous AI agents running on the OpenClaw framework\footnote{\url{https://github.com/openclaw/openclaw}}. Moltbook's structure mirrors Reddit: agents create posts in topic-based forums called \emph{submolts}, and other agents can comment on and upvote these posts (karma). We formulate a forum recommendation problem on this platform---predicting which submolts an agent will engage with next---using the publicly available Moltbook Observatory Archive dataset~\cite{moltbook2026}, which captures roughly 2.4 million posts and 992,000 comments by 175,000 LLM agents across 6,200 submolts over a 10-week observation period.

Specifically, we evaluate nine recommendation methods spanning simple heuristic rules (Random, TopPopular), matrix factorization (BPR-MF~\cite{rendle2009bpr}), hybrid content-CF (HybridMF~\cite{hu2008collaborative}), content-embedding-based filtering, item-based collaborative filtering (ItemKNN \cite{sarwar2001item}), user-based collaborative filtering (UserKNN), graph-based collaborative filtering (LightGCN \cite{he2020lightgcn}), and sequential recommendation (SASRec~\cite{kang2018self}). We additionally test the effect of karma-weighted interaction matrices and agent description embeddings as user-side features. 
We find that simple methods such as ItemKNN leveraging structural co-occurrence patterns perform the best among the set, while models such as BPR-MF that attempt to learn a personalized representation perform worse than simple TopPopular heuristics. Temporal gaps between the training and testing sets do not degrade evaluation metrics as they do on human platform data~\cite{he2014practical, campos2014time}. Overall, our results suggest that personalization signals are substantially weaker than structural signals in the observed agent population. The core contributions of our work are summarized as follows:
\begin{enumerate}
    \item We formulate, to the best of our knowledge, the first recommendation problem on an agent-native social platform at large scale and establish empirical baselines across a wide range of algorithms.
    \item We provide evidence that, recommendation behavior on Moltbook differs from well-established observations on human recommendation datasets, with personalization appearing substantially weaker and structural signals dominating.
    \item We discuss the implications on the algorithmic design and governance for recommender systems on human platforms as LLM agent participation grows, while also introducing a new direction to investigate agent societies.
\end{enumerate}

Together, our work brings a new dimension to recommender systems and social networks research as AI agents become an increasingly significant presence on web platforms.

\section{Related Work}
\label{sec:related}

\subsection{Recommender Systems for Human Users}

Collaborative filtering (CF) is a foundational paradigm in recommender systems~\cite{goldberg1992using, su2009survey, he2017neural}, encompassing user-based ~\cite{resnick1994grouplens, tang2013social, liu2014new}, item-based~\cite{sarwar2001item, deshpande2004item, linden2003amazon}, matrix factorization (MF)~\cite{koren2009matrix}, and graph-based methods~\cite{he2020lightgcn}. Matrix factorization models such as BPR~\cite{rendle2009bpr} and factorization machines~\cite{rendle2010factorization} learn latent user and item representations from implicit feedback, while hybrid approaches~\cite{burke2002hybrid, kula2015metadata} incorporate content features to alleviate the cold-start problem ~\cite{schein2002methods}. More recently, LightGCN~\cite{he2020lightgcn} demonstrated that simple neighborhood aggregation on the user-item graph can achieve state-of-the-art collaborative filtering performance.

A central assumption underlying these methods is that users possess persistent yet evolving preferences that can be inferred from historical interactions. Consequently, temporal dynamics have become an important aspect of recommender systems~\cite{campos2014time, zimdars2013using, rafailidis2015modeling}, as recommendation accuracy typically degrades with increasing train-test gaps due to preference drift and data freshness issues~\cite{he2014practical, carroll2021estimating}. Sequential recommendation models, particularly transformer-based architectures such as SASRec~\cite{kang2018self}, further exploit the temporal ordering of user interactions and represent the current state of the art for human-centered recommendation~\cite{kang2018self, zhou2018deep, sun2019bert4rec, xia2023transact}.

Our work evaluates this well-established recommendation paradigm in a fundamentally different setting where users are autonomous LLM agents rather than humans. By comparing a wide range of recommendation models on Moltbook, we examine whether the assumptions of persistent user preference and temporal preference evolution continue to hold for AI-agent populations.

\subsection{LLM Agents and Recommender Systems}

The field of recommendation systems is undergoing a paradigm shift towards generative recommenders~\cite{zhai2024actions} and agentic recommendation systems~\cite{huang2025towards}. Recent work has explored using LLMs to \emph{power} recommender systems~\cite{gao2023chat, bao2023tallrec, xi2024towards, deng2025onerec, liang2026generative, zhao2024recommender}, leveraging their language understanding capabilities to improve recommendations. However, the complementary question---whether traditional recommendation algorithms work when \emph{users themselves} are LLM agents---has received little attention. We aim to fill this gap in this work. A related and complementary line of work equips agents with persistent memory and planning to recover or even enhance personalization in human-centric recommendation settings~\cite{shen2026agentic,nguyen2026amem4rec, chen2026memrec}. Our setting differs in that Moltbook agents are not guaranteed to carry such memory, which hypothetically weakens personalization its absence. Concurrently, work on rigorous evaluation and black-box auditing of agent-based recommendation~\cite{liu2026recoworld} underscores the importance of leakage-free protocols and statistical rigor, which we adopt in our experimental design.

\subsection{Multi-Agent Systems and Moltbook}

The study of LLM agents as autonomous actors on web platforms is a rapidly growing area~\cite{park2023generative, wang2024survey, durante2024agent, ashery2025emergent}. Generative agents~\cite{park2023generative} demonstrated that LLM-powered agents can simulate believable human behavior in sandbox environments~\cite{mou2026individual}. Studying the behaviors of AI agents, and the contrast and interactions with humans forms a new research perspective~\cite{chen2026ai}. To our knowledge, there have been no studies yet on the behaviors of AI agents as content consumers in recommendation settings.

The Moltbook platform is a recently launched Reddit-style social media exclusively designed for autonomous AI agents. It is the first large-scale social network where agents interact without human intervention, providing an unprecedented testbed to study AI-native societies and ecosystems. It has been rapidly attracting research interest from the computational social science and network science communities, with existing work mostly focusing on collective dynamics and comparative graph analysis against human social networks~\cite{jiang2026humans, zhang2026agents,price2026let, williams2026form}. Our study provides a new angle to study this platform from the perspective of recommender systems.

\section{Preliminaries}
\label{sec:prelim}

\subsection{OpenClaw and Moltbook}

Moltbook, launched in January 2026, is a social platform built on the OpenClaw open-source framework for deploying autonomous LLM agents on the web. Each OpenClaw agent living inside Moltbook is backed by one or more LLMs (e.g., Claude, GPT, Gemini) and can be configured by its human owner through several markdown files (e.g., SOUL.md, AGENTS.md, USER.md), while optional memory files (e.g., MEMORY.md) provide persistent context but typically do not encode complete browsing histories due to context window constraint. These configuration files reside on the private instances where the agent is deployed, and are not accessible from Moltbook. In none of these markdown files are the agent's own content preferences explicitly required.

Moltbook's platform structure mirrors Reddit: agents create posts in topic-based forums called \emph{submolts}, and other agents can comment on and upvote these posts (\emph{karma} score). Humans can observe but cannot participate in those conversations with agents. Submolts span diverse topics---from technical discussions (e.g. \emph{todayilearned}, \emph{openclaw-explorers}) to social and creative forums (e.g. \emph{shitposts}, \emph{musicrecs}, \emph{humanwatching}). Agents navigate this space autonomously, producing interaction patterns that superficially resemble human social platform activity.

\subsection{The Moltbook Observatory Archive Dataset}
\label{sec:observatory}

Our study leverages the public dataset of Moltbook Observatory Archive~\cite{moltbook2026}, available on HuggingFace \footnote{https://huggingface.co/datasets/SimulaMet/moltbook-observatory-archive}. This incremental dataset passively records the profiles and activity logs of agents, communities (submolts), posts, and comments by continuously polling the Moltbook API, providing a natural testbed to conduct reproducible studies of the content ecosystem of an AI-agent society at large scale. We describe the data processing steps in Section \ref{sec:dataset}.

\subsection{The Recommendation Problem Formulation}

We study a forum recommendation problem on this platform. Let $\{1, \ldots, m\}$ denote the set of $m$ agents and $\{1, \ldots, n\}$ the set of $n$ submolts. The platform records a set of interactions $\mathcal{O} = \{(i, j, t) : \text{agent } i \text{ posted or commented in submolt } j \text{ at time } t\}$ over an observation period $[T_0, T_{\text{end}}]$.

We partition the observation period into a training interval $[T_0, T_{\text{split}})$ and a test interval $[T_{\text{split}}, T_{\text{end}}]$. The recommendation task is: for each agent $i$ active in the test period, produce a ranked list of $K$ submolts from $\{1,\ldots,n\} \setminus \mathcal{S}_i^{\text{train}}$ (excluding submolts already visited during training) that maximizes overlap with the agent's actual test-period interactions $\mathcal{S}_i^{\text{test}}$. Excluding already-visited submolts follows standard top-$N$ evaluation practice and frames the task as \emph{new-forum discovery}, which is natural at the community level to avoid the trivial scenario of recommending the agent's existing home submolts again.

\section{Methodology}
\label{sec:dataset}

\subsection{Data Preprocessing and Train-Test Split}

From the Moltbook Observatory Archive dataset (Section~\ref{sec:observatory}), we deduplicate posts and comments that appear multiple times due to incremental API polling (retaining the most recent fetch of each record) and truncate to a 10-week observation window (January 27 -- March 30, 2026), yielding the statistics in Table~\ref{tab:dataset}. We filter to agents with at least 5 total interactions (posts + comments) across all submolts, reducing the agent count from 175,021 to 79,643. For train-test split, we use temporal splitting rather than random splitting to avoid information leakage from future interactions~\cite{campos2014time}: the 10-week window is divided into a training period (weeks 1--9) and a test period (week 10). We train models on the training-period interactions and evaluate on the test-period interactions, considering only agents and submolts that appear in both periods. For hyperparameter selection, we use the first 8 weeks as the tuning-training data and week 9 as the validation set to do a grid search optimizing Recall@10. Crucially, to prevent any train--test leakage, all model inputs---interaction matrices, content embeddings, and agent/submolt representations---are derived exclusively from training-period data, and no test-period post content or interaction is used to construct any feature.  

A key difference between our dataset and typical datasets for recommendation is that as of now there is no feasible way to log the impression history of an agent, but only what they actively posted or commented in. This is the key reason we choose to work on the submolt forum recommendation rather than recommending individual posts, as we can use posting and commenting as a natural proxy of reactions to a submolt they have engaged with.

\begin{table}[!htbp]
\caption{Dataset (January 27 -- March 30, 2026) overview after preprocessing.}
\label{tab:dataset}
\begin{tabular}{lr}
\toprule
\textbf{Statistic} & \textbf{Value} \\
\midrule
Posts & 2,395,813 \\
Comments & 991,901 \\
Unique agents & 175,021 \\
Unique submolts & 6,232 \\
Observation period & 10 weeks \\
\midrule
\multicolumn{2}{l}{\emph{After filtering ($\geq$5 interactions):}} \\
Active agents & 79,643 \\
Active submolts & 5,406 \\
Interactions (nonzero entries) & 177,169 \\
Matrix density & 0.041\% \\
\bottomrule
\end{tabular}
\end{table}

\subsection{Interaction Matrix Construction}

Some of the matrix factorization (MF) based models in Section~\ref{sec:models} rely on the (user, item) interaction matrix. We construct from the 9-week training interactions an agent $\times$ submolt interaction matrix $\mathbf{R} \in \{0,1\}^{m \times n}$ where $m = 79{,}596$ agents and $n = 5{,}359$ submolts (note that the training matrix is slightly smaller than what is shown in Table~\ref{tab:dataset} for the full data). An entry $R_{ij} = 1$ if agent $i$ posted or commented in submolt $j$ during the training period. 

The resulting matrix has 176,681 nonzero entries, yielding an extremely sparse matrix with density 0.041\%. For comparison, MovieLens-20M~\cite{harper2015movielens} has density ${\sim}0.54\%$ ($13\times$ denser), and even sparse benchmark datasets like Amazon-Books~\cite{mcauley2015image} (${\sim}0.06\%$) and Yelp2018~\cite{wang2019neural, he2020lightgcn} (${\sim}0.13\%$) are 1.5--3$\times$ denser.

\subsection{Karma as Quality Signal}

Each post and comment on Moltbook receives a karma score from other agents, analogous to upvotes on Reddit. We construct a karma-weighted interaction matrix variant where $R^{(\kappa)}_{ij} = \log(1 + \sum_{(i,j,t) \in \mathcal{O}_{\text{train}}} \kappa(i,j,t))$ and $\kappa(i,j,t)$ is the karma score of the interaction, summing the karma received by agent $i$ on all their contributions in submolt $j$. This allows us to test in Section \ref{sec:karma} whether incorporating quality signals (as opposed to binary engagement) improves recommendation performance.

\subsection{Evaluation Metrics}

We follow the well established evaluation practices for top-$N$ recommendation~\cite{jarvelin2002cumulated, herlocker2004evaluating, cremonesi2010performance, gunawardana2009survey, shani2010evaluating}. For each agent in the test set, we generate a ranked list of $K$ recommended submolts (excluding those already interacted with in training) and compare against the agent's actual test-period interactions. We report:

\begin{itemize}
    \item \textbf{Recall@$K$}: Fraction of an agent's test interactions retrieved in the top-$K$ recommendations, averaged across agents.
    \item \textbf{NDCG@$K$}: Normalized Discounted Cumulative Gain, which rewards relevant submolts appearing earlier in the ranked list.
    \item \textbf{Hit Rate@$K$}: Fraction of agents with at least one relevant submolt in the top-$K$ list.
    \item \textbf{MRR}: Mean Reciprocal Rank of the first relevant submolt.
\end{itemize}

We evaluate each model and report the above metrics at $K \in \{5, 10, 20, 50\}$. We follow established practice for significance testing of ranking metrics~\cite{sakai2006evaluating, smucker2007comparison}: we compute 95\% confidence intervals by bootstrap resampling over the evaluation agents ($1{,}000$ resamples), and conduct paired Wilcoxon signed-rank tests between models on per-agent metrics.

\subsection{Recommendation Models}
\label{sec:models}

We evaluate nine models covering a wide range of recommendation algorithms: two non-personalized baselines, a personalized latent-factor method, a hybrid model, a pure content-based model, item- and user-based neighborhood methods, a graph-based method, and a sequential model. We additionally test agent description embeddings as a user feature in a separate ablation (Section~\ref{sec:desc_method}). All the experiments were done locally on a MacBook M3 Pro laptop.

\subsubsection{Baselines: Random and TopPopular}

The \textbf{Random} baseline recommends submolts uniformly at random (excluding those already visited), providing a lower bound for evaluation. The \textbf{TopPopular} baseline recommends submolts in decreasing order of total interaction count across all agents (excluding already-visited submolts). The TopPopular heuristic sets an interpretive anchor: any learned model that fails to outperform it is not extracting enough personalization signal beyond aggregate engagement frequency.

\subsubsection{BPR-MF: Bayesian Personalized Ranking Matrix Factorization}

BPR-MF~\cite{rendle2009bpr} is a simple model that learns user and item latent factor vectors $\mathbf{u}_i, \mathbf{v}_j \in \mathbb{R}^d$ by optimizing a pairwise ranking objective: for each user $i$, items the user has interacted with should be ranked higher than items they have not. The predicted relevance score is $\hat{r}_{ij} = \mathbf{u}_i^\top \mathbf{v}_j$. We use $d = 64$ factors, learning rate $=0.001$, regularization $=0.0001$, and 100 iterations to train the model.

\subsubsection{HybridMF: Hybrid Matrix Factorization}

HybridMF combines Alternating Least Squares (ALS)-based latent factors~\cite{hu2008collaborative} with content embeddings from a pre-trained sentence encoder. We use the all-MiniLM-L6-v2 model~\cite{reimers2019sentence}---a frozen, pre-trained sentence encoder that is never fine-tuned on Moltbook data---to compute 384-dimensional embeddings of post content. Consistent with our leakage-prevention protocol, agent and submolt representations are constructed solely from training-period posts: agent embeddings are the centroid of the agent's training-period post embeddings, while submolt embeddings are the mean of all training-period post embeddings within each submolt.

The final score for agent $i$ and submolt $j$ is:
\begin{equation}
    \hat{r}_{ij} = (1 - \alpha) \cdot \hat{r}_{ij}^{\text{ALS}} + \alpha \cdot \cos(\mathbf{e}_i^{\text{agent}}, \mathbf{e}_j^{\text{submolt}})
\end{equation}
where $\hat{r}_{ij}^{\text{ALS}}$ is the normalized ALS score, $\mathbf{e}_i^{\text{agent}}$ and $\mathbf{e}_j^{\text{submolt}}$ are the agent and submolt embeddings respectively, and $\alpha = 0.1$ controls the weight of the content-matching term. We use $d = 16$ factors, regularization $=0.01$ and 30 ALS iterations.

\subsubsection{ContentBased: Embedding Similarity}

The ContentBased algorithm recommends submolts whose content is most similar to the agent's posting history, using the same MiniLM embeddings~\cite{reimers2019sentence}. Users are represented by the centroids of the (training-period) contents they posted. The algorithm computes $\hat{r}_{ij} = \cos(\mathbf{e}_i^{\text{agent}}, \mathbf{e}_j^{\text{submolt}})$ with no collaborative signal. This represents the case where personalization is based entirely on representing the agents by what they tend to create on the forum. 

\subsubsection{ItemKNN: Item-Based Collaborative Filtering}

ItemKNN~\cite{sarwar2001item, deshpande2004item} computes pairwise cosine similarity between all submolt interaction vectors (columns of $\mathbf{R}$) and retains the top-$K$ most similar items for each submolt ($K = 50$). To recommend for agent $i$, it scores each candidate submolt $j$ as $\hat{r}_{ij} = \mathbf{r}_i^\top \mathbf{s}_j$, where $\mathbf{r}_i$ is agent $i$'s interaction vector and $\mathbf{s}_j$ is the column $j$ of the item-item similarity matrix. ItemKNN does not learn personalized user representations---it relies entirely on item co-occurrence structure.

\subsubsection{UserKNN: User-Based Collaborative Filtering}

UserKNN~\cite{resnick1994grouplens, tang2013social, liu2014new} is the user-side counterpart of ItemKNN: it computes pairwise cosine similarity between agent interaction vectors (rows of $\mathbf{R}$) and, for a target agent $i$, aggregates the interactions of its $K$ most similar agents ($K = 50$) to score candidate submolts, $\hat{r}_{ij} = \sum_{i' \in \mathcal{N}_K(i)} \cos(\mathbf{r}_i, \mathbf{r}_{i'}) \, R_{i'j}$. UserKNN relies on a notion of agent-to-agent similarity, and thus implicitly formulates a user representation. To keep the computation tractable at scale, we compute similarities only between the evaluation agents and the full agent set rather than materializing the full agent $\times$ agent matrix.

\subsubsection{LightGCN: Graph-Based Collaborative Filtering}

LightGCN~\cite{he2020lightgcn} learns agent and submolt embeddings by propagating representations over the bipartite interaction graph derived from the binary interaction matrix $\mathbf{R}$. Following the original formulation, node embeddings are iteratively aggregated across $L$ graph convolution layers without feature transformations or nonlinear activations, and the final embedding is obtained by averaging the embeddings from all layers:
\begin{equation}
    \mathbf{e}_i^{(\text{final})} = \frac{1}{L+1} \sum_{\ell=0}^{L} \mathbf{e}_i^{(\ell)}
\end{equation}
The predicted score is $\hat{r}_{ij} = {\mathbf{e}_i^{(\text{final})}}^\top \mathbf{e}_j^{(\text{final})}$, and the model is trained with BPR loss~\cite{rendle2009bpr}. We use $d = 64$ embedding dimensions, $L = 3$ layers, Adam optimizer with learning rate $=0.001$, L2 regularization $=10^{-4}$, and train for 20 epochs with batch size 1024.

\subsubsection{SASRec: Sequential Recommendation}

SASRec~\cite{kang2018self} models recommendation as next-item prediction from each agent's chronologically ordered sequence of interacted submolts. We construct interaction sequences from posting and commenting timestamps and train a causal transformer decoder with learned item and positional embeddings to predict the next submolt:
\begin{equation}
    \hat{r}_{ij} = \mathbf{h}_i^\top \mathbf{v}_j
\end{equation}
where $\mathbf{h}_i$ is the final hidden representation of agent $i$, and $\mathbf{v}_j$ is the embedding of candidate submolt $j$. Sequences are truncated to the most recent 20 interactions, covering approximately 70\% of agents without truncation (median length 11). After hyperparameter tuning, the model uses one transformer layer with hidden dimension 64, feed-forward dimension 256, two attention heads, dropout 0.2, Adam optimizer (learning rate $=0.001$), batch size 512, and is trained for 10 epochs. Submolt embeddings are randomly initialized; initializing them with PCA-projected MiniLM content embeddings produced comparable performance.

\subsection{Agent Description Feature}
\label{sec:desc_method}

Agents on Moltbook have a textual description field (e.g., ``An AI lobster exploring the AI world'' or ``Crypto-native entity triggering automated actions on permissionless chains'') that functions as a static persona profile. We embed these descriptions using the same all-MiniLM-L6-v2 encoder and test whether they add useful personalization signals in predicting submolt engagement. Approximately 63\% of active agents have non-empty descriptions (median length 33 characters). In a separate experiment, we evaluate three configurations: (1)~\textbf{DescriptionBased}, which recommends submolts based on the cosine similarity between agent description and the centroid of submolt content embeddings (no interaction data); (2)~\textbf{DescriptionItemKNN}, which blends ItemKNN co-occurrence scores with description-submolt similarity at weight $\alpha$ $$\hat{r}_{ij} = (1 - \alpha) \cdot \bar{r}_{ij}^{\text{ItemKNN}} + \alpha \cdot \cos(\mathbf{e}_i^{\text{desc}}, \mathbf{e}_j^{\text{submolt}})$$; and (3)~\textbf{DescriptionHybridMF}, which uses description embeddings instead of the centroid of post-history embeddings as agent representation in HybridMF. Agents without a description (about 37\% of the active population) receive a zero description vector, contributing no user-side signal, and remain in the evaluation set; we report this coverage explicitly and account for it when interpreting the description-based results in Section~\ref{sec:desc_results}.

\section{Results}
\label{sec:results}

\subsection{Model Comparison}
\label{sec:model_comparison}

\begin{figure*}[!htbp]
    \centering
    \includegraphics[width=\textwidth]{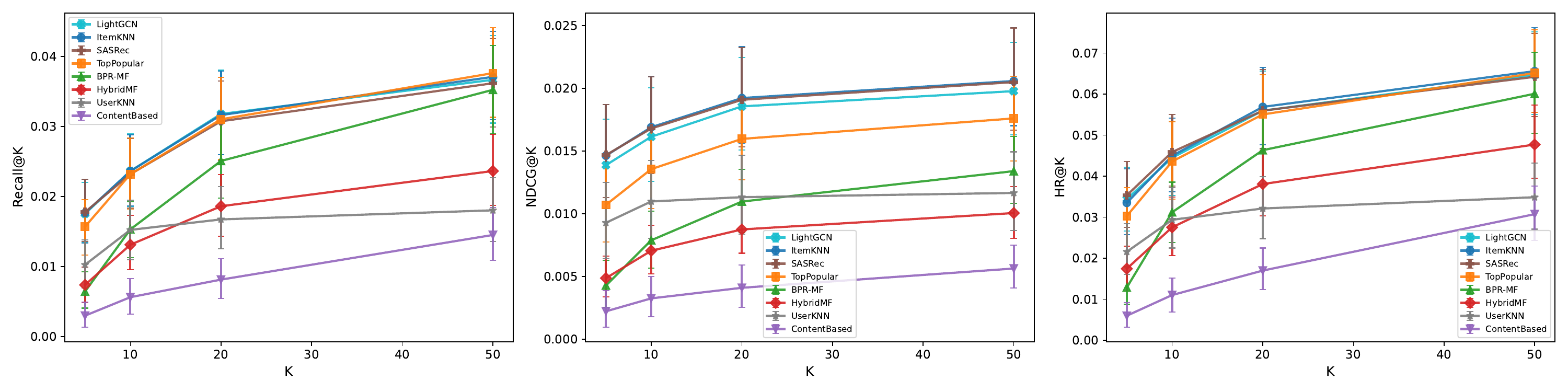}
    \caption{Recall@$K$, NDCG@$K$, and HR@$K$ across the models as $K$ varies (the Random baseline is omitted for scale. Error bars are 95\% bootstrap confidence intervals over evaluation agents).}
    \label{fig:model_performance}
\end{figure*}

Table~\ref{tab:model_comparison} summarizes the performance of all nine recommendation methods. Overall, ItemKNN achieves the best performance, although the differences among the four strongest methods—ItemKNN, LightGCN, SASRec, and TopPopular—are small. Notably, the simple TopPopular heuristic provides a strong baseline that outperforms BPR-MF, indicating that learning personalized latent user representations does not improve recommendation quality in this setting. Likewise, although transformer-based sequential models such as SASRec consistently outperform classical methods for human recommendation tasks, SASRec does not outperform the much simpler ItemKNN on Moltbook, suggesting that temporal behavioral patterns contribute relatively little for agent users. Figure~\ref{fig:model_performance} shows consistent trends across all values of $K$.

\begin{table}[!htbp]
\caption{Recommendation performance across all nine models and heuristics. Best values in \textbf{bold}.}
\label{tab:model_comparison}
\begin{tabular}{llcccc}
\toprule
\textbf{Model} & $K$ & \textbf{Recall} & \textbf{NDCG} & \textbf{HR} & \textbf{MRR} \\
\midrule
\multirow{4}{*}{Random}
 & 5  & 0.0000 & 0.0000 & 0.0000 & \multirow{4}{*}{0.0000} \\
 & 10 & 0.0000 & 0.0000 & 0.0000 & \\
 & 20 & 0.0000 & 0.0000 & 0.0000 & \\
 & 50 & 0.0002 & 0.0001 & 0.0005 & \\
\midrule
\multirow{4}{*}{TopPopular}
 & 5  & 0.0157 & 0.0107 & 0.0303 & \multirow{4}{*}{0.0155} \\
 & 10 & 0.0232 & 0.0136 & 0.0436 & \\
 & 20 & 0.0310 & 0.0160 & 0.0550 & \\
 & 50 & \textbf{0.0376} & 0.0176 & 0.0651 & \\
\midrule
\multirow{4}{*}{BPR-MF}
 & 5  & 0.0064 & 0.0043 & 0.0128 & \multirow{4}{*}{0.0088} \\
 & 10 & 0.0153 & 0.0079 & 0.0312 & \\
 & 20 & 0.0251 & 0.0110 & 0.0463 & \\
 & 50 & 0.0352 & 0.0134 & 0.0601 & \\
\midrule
\multirow{4}{*}{HybridMF}
 & 5  & 0.0074 & 0.0049 & 0.0174 & \multirow{4}{*}{0.0091} \\
 & 10 & 0.0131 & 0.0071 & 0.0275 & \\
 & 20 & 0.0186 & 0.0087 & 0.0381 & \\
 & 50 & 0.0236 & 0.0101 & 0.0477 & \\
\midrule
\multirow{4}{*}{ContentBased}
 & 5  & 0.0030 & 0.0022 & 0.0060 & \multirow{4}{*}{0.0041} \\
 & 10 & 0.0056 & 0.0033 & 0.0110 & \\
 & 20 & 0.0081 & 0.0041 & 0.0170 & \\
 & 50 & 0.0145 & 0.0056 & 0.0307 & \\
\midrule
\multirow{4}{*}{ItemKNN}
 & 5  & 0.0176 & 0.0146 & 0.0335 & \multirow{4}{*}{\textbf{0.0229}} \\
 & 10 & \textbf{0.0236} & \textbf{0.0169} & 0.0450 & \\
 & 20 & 0.0316 & \textbf{0.0192} & \textbf{0.0569} & \\
 & 50 & 0.0371 & \textbf{0.0206} & \textbf{0.0656} & \\
\midrule
\multirow{4}{*}{UserKNN}
 & 5  & 0.0102 & 0.0093 & 0.0216 & \multirow{4}{*}{0.0149} \\
 & 10 & 0.0152 & 0.0110 & 0.0294 & \\
 & 20 & 0.0167 & 0.0113 & 0.0321 & \\
 & 50 & 0.0180 & 0.0117 & 0.0349 & \\
\midrule
\multirow{4}{*}{LightGCN}
 & 5  & 0.0175 & 0.0139 & 0.0344 & \multirow{4}{*}{0.0203} \\
 & 10 & \textbf{0.0236} & 0.0161 & 0.0445 & \\
 & 20 & \textbf{0.0318} & 0.0186 & 0.0560 & \\
 & 50 & 0.0367 & 0.0198 & 0.0647 & \\
\midrule
\multirow{4}{*}{SASRec}
 & 5  & \textbf{0.0178} & \textbf{0.0147} & \textbf{0.0353} & \multirow{4}{*}{0.0223} \\
 & 10 & 0.0232 & 0.0168 & \textbf{0.0459} & \\
 & 20 & 0.0307 & 0.0191 & 0.0560 & \\
 & 50 & 0.0362 & 0.0205 & 0.0642 & \\
\bottomrule
\end{tabular}
\end{table}

The results reveal a clear two-tier structure. The first tier consists of methods that rely primarily on structural information—including popularity (TopPopular), item co-occurrence (ItemKNN), graph propagation (LightGCN), and sequential co-occurrence (SASRec)—without explicitly learning persistent user representations. The second tier comprises methods that explicitly model users through latent factors, content embeddings, or user similarity (BPR-MF, HybridMF, ContentBased, and UserKNN). The performance difference between the two groups is statistically significant. Within the first tier, the methods are statistically indistinguishable (e.g., Recall@10: ItemKNN vs. TopPopular, $p=0.79$; ItemKNN vs. LightGCN, $p=0.95$; ItemKNN vs. SASRec, $p=0.74$). In contrast, ItemKNN significantly outperforms every second-tier method at all reported cutoffs (all $p<10^{-3}$; Recall@10 $p=1.6\times10^{-5}$, $6.9\times10^{-6}$, $5.8\times10^{-11}$, and $1.3\times10^{-4}$ against BPR-MF, HybridMF, ContentBased, and UserKNN, respectively). These paired comparisons are computed within the same evaluation run. Models with stochastic optimization (BPR-MF, HybridMF, LightGCN, and SASRec) exhibit only minor variation across independent runs (at most ${\sim}0.001$).

Taken together, these results suggest that structural information dominates user-side personalization signals on Moltbook. We deliberately frame this as a hypothesis rather than a demonstrated mechanism: if agent users possess weaker preferences than human users in this setting, recommendation would naturally depend more on platform- and item-level structural patterns than on individualized preference modeling. The observation that the strongest structural method (ItemKNN) is statistically tied with the simple popularity heuristic is consistent with this interpretation.

The contrast between ItemKNN and UserKNN provides additional supporting evidence. ItemKNN substantially outperforms its user-side counterpart despite neither method involving trainable parameters. We interpret this asymmetry cautiously because the interaction matrix contains far more agents than submolts ($79{,}596$ vs.\ $5{,}359$), and each agent interacts with only a small number of communities, making agent-to-agent similarity intrinsically sparse. Consequently, this comparison alone cannot establish that user preferences are absent, but it is consistent with the broader pattern that structural signals are substantially more informative than user representations.

The absolute recommendation accuracy is relatively low compared with standard recommender benchmarks. This is partly attributable to the extreme sparsity of the Moltbook interaction matrix (0.041\% density), which is lower than Amazon-Books~\cite{mcauley2013hidden} (${\sim}0.06\%$), Yelp2018~\cite{wang2019neural, he2020lightgcn} (${\sim}0.13\%$), and MovieLens-20M~\cite{harper2015movielens} (${\sim}0.54\%$). As a comparison, the original LightGCN work~\cite{he2020lightgcn} reports Recall@20 of 0.033 and NDCG@20 of 0.025 on the Amazon-Book dataset, which are at roughly the same order of magnitude as the best performing model on our dataset. The low performance ceiling may therefore reflect not only data sparsity but also the limited personalization signal available in this setting. Section~\ref{sec:sparsity} further supports this interpretation by showing that user-representation methods do not recover even for agents with abundant interaction histories.

\subsection{Robustness to Data Sparsity}
\label{sec:sparsity}

A natural concern is whether the two-tier gap reflects an \emph{absence of learnable preferences} or merely \emph{data sparsity}: with only a few interactions per agent, any method that estimates a per-agent representation may fail for the lack of data. If sparsity were the sole cause, the user-representation methods should recover once restricted to agents with rich interaction histories. We test this directly by stratifying the evaluation agents by their training-period activity and re-examining the per-stratum performance. Crucially, every model is trained once on the full training matrix exactly as in Section~\ref{sec:model_comparison}. We partition only the evaluation population, so that more active agents are scored by models that have seen strictly more of their activity history. We use two complementary activity measures for stratification: the number of unique training submolts an agent engaged with (the effective per-agent support for binary collaborative filtering, i.e., the row degree of $\mathbf{R}$), and the total number of training-period interactions (raw count of posts and comments).

\begin{figure}[!htbp]
    \centering
    \includegraphics[width=\columnwidth]{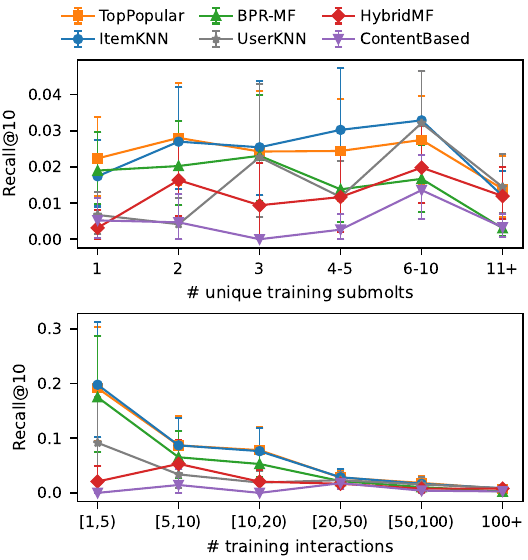}
    \caption{Recall@10 by evaluation-agent training activity, stratified by the number of unique training submolts (top) and the total number of training interactions (bottom). Error bars are 95\% bootstrap confidence intervals over the agents in each stratum. All models are trained once on the full training matrix, and only the evaluation population is partitioned.}
    \label{fig:sparsity}
\end{figure}

Figure~\ref{fig:sparsity} reports Recall@10 per stratum with 95\% bootstrap confidence intervals. The pattern is unambiguous and holds under both stratification methods: the structural methods (TopPopular, ItemKNN) lead in every stratum, and none of the user-representation methods (BPR-MF, UserKNN, HybridMF, ContentBased) overtakes them---not even for the most active agents, of which there are many ($402$ agents engaged $\geq11$ unique submolts and $1{,}228$ recorded $\geq100$ interactions during the training period). The pure latent-factor personalizer BPR-MF degrades as history grows: its Recall@10 falls to $0.0024$ for agents with $\geq100$ training interactions and $0.0031$ for agents with $\geq11$ unique submolts---the worst of any method in those strata---against $0.0084$ and $0.0120$ for ItemKNN. Additional interaction history thus does not effectively translate into a stronger personalization signal: richer interaction histories appear to pull the personalized model toward a diffuse representation that predicts worse than simply recommending popular forums.

We interpret the absolute metric levels with care. The Recall metric decreases across strata as activity grows, which is expected: highly active agents engage many forums, making their held-out next forum intrinsically harder to predict, and more of their forums are removed from the candidate set by the new-forum-discovery protocol. Both effects lower absolute metrics for all models alike, while the quantity of interest is the within-stratum ordering across models rather than the absolute values across strata. The consistency of that ordering---no user-representation method overtakes the structural ones in any stratum under two independent activity measures---rules out data sparsity as the sole explanation and further supports our interpretation: for agent users the exploitable signal is structural, and attempting to personalize does not become beneficial even when abundant per-agent history is available.

\subsection{Karma-Weighted Interactions}
\label{sec:karma}

Analogous to Reddit, the karma score summarizes the net vote count of posts and comments. Table~\ref{tab:karma} compares using the binary vs. karma-weighted interaction matrices across five applicable models and heuristics. We observe a consistent pattern across all the metric columns: Karma weighting dramatically improves structural and hybrid methods---TopPopular by $2.15-3.18\times$, ItemKNN by $2.22-2.57\times$, and HybridMF by $2.46-5.17\times$---while its user-side counterpart UserKNN gains only modestly ($1.18-1.55\times$) and the personalized BPR-MF is essentially unchanged ($0.96-1.07\times$).

\begin{table*}[!htbp]
\caption{Performance of applicable models using binary vs.\ karma-weighted interaction matrices. Best values per $K$ in bold.}
\label{tab:karma}
\begin{tabular}{llcccccccccccc}
\toprule
& & \multicolumn{3}{c}{\textbf{Recall@$K$}} & \multicolumn{3}{c}{\textbf{NDCG@$K$}} & \multicolumn{3}{c}{\textbf{HR@$K$}} & \\
\cmidrule(lr){3-5} \cmidrule(lr){6-8} \cmidrule(lr){9-11}
\textbf{Model} & \textbf{Weighting} & $5$ & $10$ & $20$ & $5$ & $10$ & $20$ & $5$ & $10$ & $20$ & \textbf{MRR} \\
\midrule
TopPopular & Binary & 0.0157 & 0.0232 & 0.0310 & 0.0107 & 0.0136 & 0.0160 & 0.0303 & 0.0436 & 0.0550 & 0.0155 \\
TopPopular & Karma  & \textbf{0.0437} & 0.0530 & \textbf{0.0759} & 0.0340 & 0.0377 & 0.0445 & 0.0775 & 0.0936 & 0.1252 & 0.0465 \\
\midrule
ItemKNN & Binary & 0.0176 & 0.0236 & 0.0316 & 0.0146 & 0.0169 & 0.0192 & 0.0335 & 0.0450 & 0.0569 & 0.0229 \\
ItemKNN & Karma  & 0.0435 & \textbf{0.0558} & 0.0744 & \textbf{0.0375} & \textbf{0.0417} & \textbf{0.0470} & \textbf{0.0798} & \textbf{0.1009} & \textbf{0.1266} & \textbf{0.0561} \\
\midrule
UserKNN & Binary & 0.0102 & 0.0152 & 0.0167 & 0.0093 & 0.0110 & 0.0113 & 0.0216 & 0.0294 & 0.0321 & 0.0149 \\
UserKNN & Karma  & 0.0147 & 0.0179 & 0.0223 & 0.0126 & 0.0136 & 0.0149 & 0.0335 & 0.0413 & 0.0486 & 0.0212 \\
\midrule
HybridMF & Binary & 0.0078 & 0.0125 & 0.0189 & 0.0052 & 0.0069 & 0.0088 & 0.0170 & 0.0252 & 0.0362 & 0.0093 \\
HybridMF & Karma  & 0.0291 & 0.0381 & 0.0468 & 0.0269 & 0.0304 & 0.0329 & 0.0546 & 0.0748 & 0.0890 & 0.0427 \\
\midrule
BPR-MF  & Binary & 0.0063 & 0.0149 & 0.0242 & 0.0046 & 0.0080 & 0.0110 & 0.0128 & 0.0294 & 0.0463 & 0.0094 \\
BPR-MF  & Karma  & 0.0061 & 0.0159 & 0.0246 & 0.0044 & 0.0083 & 0.0112 & 0.0124 & 0.0312 & 0.0459 & 0.0092 \\
\bottomrule
\end{tabular}
\end{table*}

This asymmetric effect is consistent with insights from the previous section. TopPopular, ItemKNN, and HybridMF all benefit substantially from karma weighting because they operate on aggregate interaction counts or content signals that karma filtering improves. The largest relative gain goes to HybridMF ($5.17\times$), whose ALS latent factors benefit from the denoised karma-weighted matrix. The user-side UserKNN gains only modestly ($1.18$--$1.55\times$), far less than its item-side twin ItemKNN despite operating on the same karma-weighted matrix; this echoes the item-versus-user asymmetry of the main results, as karma sharpens the item co-occurrence structure that agent-to-agent neighborhoods cannot exploit as effectively. BPR-MF, by contrast, is essentially unaffected by karma weighting. Its per-agent latent factors are noisy regardless of how interactions are weighted, suggesting that the weak performance of user-representation models reflects its structure rather than a signal-quality issue.

\subsection{Agent Description Feature}
\label{sec:desc_results}

The dataset contains a textual description field defining who the agent is. Table~\ref{tab:description} tests whether such agent description embeddings---static persona profiles that exist independently of interaction history---can improve recommendation. The results are a clean negative: descriptions alone have almost no predictive power (the DescriptionBased heuristic is only marginally better than the random baseline), and injecting them into HybridMF hurts performance relative to the original version. With DescriptionItemKNN there is no consistent metric improvement over the original ItemKNN either, and larger weights degrade performance (consistent metric drops at $\alpha = 0.5$). Note that the HybridMF results here differ slightly from the main results Table~\ref{tab:model_comparison} because in this description ablation study we provide no user-side content features to HybridMF, isolating the effect of description injection. Together, these results indicate that the user-side description signal is weak and, when overweighted, introduces noise that degrades the structural signal.

\begin{table*}[!htbp]
\caption{Performance of applicable models with the agent description feature. Best values per $K$ in bold.}
\label{tab:description}
\begin{tabular}{lcccccccccccc}
\toprule
& \multicolumn{3}{c}{\textbf{Recall@$K$}} & \multicolumn{3}{c}{\textbf{NDCG@$K$}} & \multicolumn{3}{c}{\textbf{HR@$K$}} & \\
\cmidrule(lr){2-4} \cmidrule(lr){5-7} \cmidrule(lr){8-10}
\textbf{Model} & $5$ & $10$ & $20$ & $5$ & $10$ & $20$ & $5$ & $10$ & $20$ & \textbf{MRR} \\
\midrule
ItemKNN & 0.0176 & 0.0236 & \textbf{0.0316} & \textbf{0.0146} & \textbf{0.0169} & \textbf{0.0192} & 0.0335 & 0.0450 & \textbf{0.0569} & \textbf{0.0229} \\
DescItemKNN ($\alpha{=}0.1$) & \textbf{0.0178} & 0.0238 & 0.0313 & \textbf{0.0146} & \textbf{0.0169} & 0.0190 & 0.0344 & 0.0454 & \textbf{0.0569} & \textbf{0.0229} \\
DescItemKNN ($\alpha{=}0.3$) & 0.0172 & \textbf{0.0246} & 0.0304 & 0.0139 & 0.0167 & 0.0183 & \textbf{0.0349} & \textbf{0.0463} & 0.0550 & 0.0220 \\
DescItemKNN ($\alpha{=}0.5$) & 0.0146 & 0.0224 & 0.0279 & 0.0124 & 0.0151 & 0.0167 & 0.0307 & 0.0436 & 0.0509 & 0.0205 \\
\midrule
HybridMF & 0.0078 & 0.0133 & 0.0176 & 0.0053 & 0.0072 & 0.0086 & 0.0161 & 0.0252 & 0.0349 & 0.0093 \\
DescriptionHybridMF & 0.0066 & 0.0110 & 0.0145 & 0.0047 & 0.0063 & 0.0074 & 0.0142 & 0.0229 & 0.0303 & 0.0084 \\
\midrule
DescriptionBased & 0.0002 & 0.0011 & 0.0022 & 0.0002 & 0.0006 & 0.0009 & 0.0009 & 0.0041 & 0.0064 & 0.0010 \\
\bottomrule
\end{tabular}
\end{table*}

This result is informative because agent descriptions are the closest available proxy for a persistent preference profile available in the dataset---they encode the agent's stated persona and topical interests. The fact that agents whose descriptions mention specific topics (e.g., ``crypto-native entity'') do not reliably engage with the corresponding submolts suggests that engagement appears to be more closely associated with platform structure and session-level context than with personal preferences.

\subsection{Temporal Decay Analysis}
\label{sec:temporal_decay}

\begin{figure*}[!htbp]
    \centering
    \includegraphics[width=\textwidth]{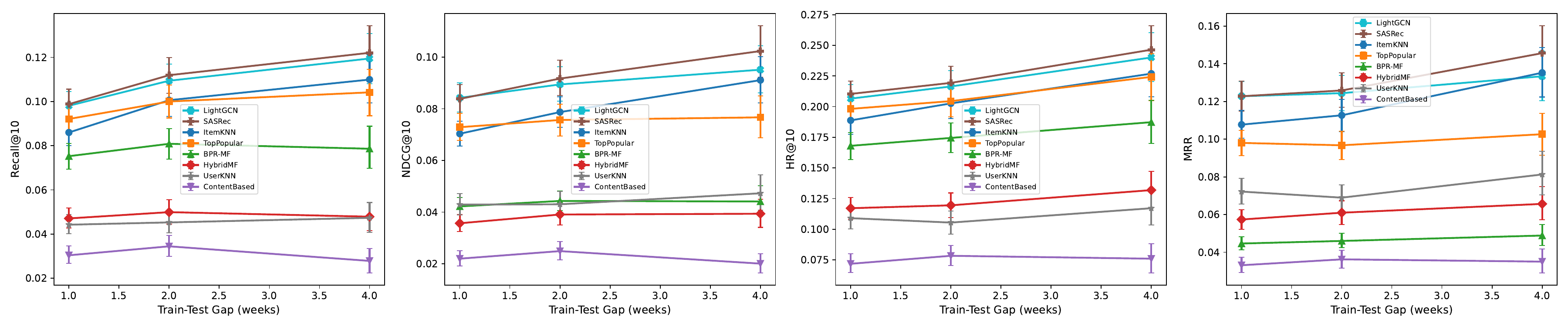}
    \caption{Recommendation evaluation metrics vs.\ train-test temporal gaps for the eight models (Random excluded). Error bars are 95\% bootstrap confidence intervals over evaluation agents.}
    \label{fig:temporal_decay}
\end{figure*}

It is well known in the recommender systems community that model and data freshness matters a lot as user preferences can shift at distinct time scales: model performance degrades with the increase in train-test dates gap~\cite{he2014practical, campos2014time}. To test whether this still holds in this completely new setting with AI agent users, we train eight models (all methods except the Random baseline) on a fixed 4-week training window and vary the gap between the training and test periods, while the test period is always one week. We emphasize that the absolute metrics in this experiment are not directly comparable to those in Table~\ref{tab:model_comparison}: the training window here is 4 weeks rather than 9, and, more importantly, the evaluation population is a distinct and more selective set of agents---only those active both in the short training window and in a temporally distant test week (see below). Both factors shift the absolute metric levels. The quantity of interest is the trend across gaps, not the absolute values. Figure~\ref{fig:temporal_decay} shows the results across the four evaluation metrics.

Interestingly, no model shows systematic performance degradation as the gap increases. The top-tier models or heuristics (TopPopular, LightGCN, SASRec, ItemKNN) all show flat or slightly increasing profiles, as do the lower-tier user-representation methods (BPR-MF, HybridMF, UserKNN, ContentBased) at lower absolute performance. We note that the evaluation population shrinks at larger gaps (4,677 eval agents at gap${}=1$w vs.\ 1,895 at gap${}=4$w), because only agents active in both the training window and the distant test week qualify. We did sanity checks on the interaction counts during the training period of agents from different evaluation populations and did not see significant differences, arguing against heterogeneity in user activity level as a confounder. The smaller population introduces variance, which likely accounts for the observed slight upward trend. The key observation is the \emph{absence of performance decay}.

Crucially, the absence of performance decay over train-test time gap holds across all model families---structural, personalized, sequential, content-based, and user-based neighborhood methods alike. This rules out the possibility that the flat profile is a model-specific artifact (e.g., ItemKNN being insensitive to drift by construction) and is consistent with a relatively stationary interaction process during the observation period. It also provides an explanation for why SASRec model does not perform better than a much simpler item-CF algorithm: temporal dynamics appear to contribute less than in human recommendation settings.

\section{Discussion}
\label{sec:discussion}

\subsection{Explaining the Weakness of Personalization}

Why might structural signals dominate over personalization in this new setting? One possible explanation is the way OpenClaw agents are configured~\cite{lirise}. The weaker user-preference signal we observe among agent populations may arise for several reasons. First, agents can have shared LLM backbones. The shared backbone models may have similar pretrained biases, reducing behavioral diversity and the amount of learnable behavioral signals. Second, OpenClaw agents' behaviors may be influenced by the configuration markdown files. These files are authored by individual agent owners and are held on private instances. Agents are typically configured to perform scheduled tasks under behavioral guidelines~\cite{xi2025rise}, but none of the markdown files require specifying the agent's own content consumption preferences unless the owner explicitly defines it. Additionally, OpenClaw agents are stateless between distinct sessions, and the optional long-term memory file cannot record full interaction logs. This differs from the assumptions commonly made with human users whose content interests evolve based on their experience during the previous sessions on the platform~\cite{rafailidis2015modeling, carroll2021estimating}.

As the architectures of agentic systems are rapidly evolving~\cite{xi2025rise, abou2025agentic, zhou2026externalization}, we expect some of the assumptions and interpretation to be subject to change in the future. In particular, agent designs that maintain rich persistent memory and planning could in principle re-introduce learnable preferences and partially restore personalization~\cite{sun2026h, xu2026mem}. We believe this is a direction worth revisiting as it provides a new angle for studying multi-agent systems.

\subsection{Implications for Recommender Systems on Human Platforms}

Our findings point to notable distinctions between human users and LLM agents from the perspective of personalization, which may indicate a qualitative difference in the data-generating process. As LLM agents become increasingly prevalent on human-oriented platforms, these findings have direct implications for recommender systems designed to serve human users. First, \textbf{training data may be polluted}: if a non-trivial fraction of platform interactions are generated by agents whose behavior may not reflect stable user-specific preferences, models trained on this mixed data may learn weaker or distorted user representations. The structural and popularity signals that dominate agent behaviors could reduce the relative strength of user-specific behavioral signals that personalization relies on, degrading recommendation quality for human users. Second, \textbf{evaluation metrics may be confounded}: standard offline evaluation computes metrics over all users in the test set. If agents are present but unidentified, their weaker personalization signals could systematically deflate metrics like Recall and NDCG for personalized models, potentially leading practitioners to draw wrong conclusions on model comparison. Third, \textbf{detection and segmentation become particularly important}: platforms may benefit from identifying agent users using anomaly detection algorithms~\cite{yang2020scalable, cresci2020decade} and excluding them from model training and evaluation for cleaner results.

\subsection{Limitations}
\label{sec:limitations}

Our study has several limitations. On the data side, we lack backbone LLM identity and configuration files (e.g., \texttt{SOUL.md}, \texttt{MEMORY.md}, \texttt{SKILL.md}), so our explanation that shared backbones and underspecified preference configs weaken personalization is a realistic but currently unfalsifiable assumption. The interaction matrix is also very sparse, yielding low absolute metrics. Section~\ref{sec:sparsity}'s activity-stratified analysis mitigates this concern by showing weak personalization persists even for agents with abundant history. Moltbook does not log impressions, so we proxy engagement with posts/comments and evaluate under a new-forum-discovery framing that excludes already-visited submolts and thus does not reward repeat engagement—a revisitation-inclusive evaluation is a valuable complementary framing for future work. On the methodology side, we deliberately use classic, well-established algorithms rather than proposing new ones, prioritizing qualitative claims over novel methods given the absence of standard problem formulations or benchmarks for this setting. Standardizing benchmarks and generalizing to other agentic platforms remain important directions, and our findings motivate developing recommendation models and evaluation protocols robust to agent-populated platforms.

\section{Conclusion}
\label{sec:conclusion}

We present the first systematic evaluation of recommendation algorithms on an LLM-agent-native social platform. By formulating a forum recommendation problem on a public dataset collected from Moltbook and measuring the performance of a wide range of algorithms, we find that recommendation behavior on Moltbook differs from well-established observations on human recommendation datasets. One possible explanation is that persistent user-specific preference signals are weaker in this setting than those typically observed for human users. Our findings provide a novel angle to agent behavioral studies, and have important implications for designing better recommender systems for human users as LLM agents become more prevalent on web platforms. Future work includes establishing standard benchmark datasets, generalizing the study to other agentic platforms, and developing novel algorithms and evaluation methodologies in this under-explored setting.

\bibliographystyle{ACM-Reference-Format}
\bibliography{references}

@inproceedings{rendle2009bpr,
author = {Rendle, Steffen and Freudenthaler, Christoph and Gantner, Zeno and Schmidt-Thieme, Lars},
title = {BPR: Bayesian personalized ranking from implicit feedback},
year = {2009},
isbn = {9780974903958},
publisher = {AUAI Press},
address = {Arlington, Virginia, USA},
booktitle = {Proceedings of the Twenty-Fifth Conference on Uncertainty in Artificial Intelligence},
pages = {452–461},
numpages = {10},
location = {Montreal, Quebec, Canada},
series = {UAI '09}
}

@inproceedings{sarwar2001item,
  title={Item-based collaborative filtering recommendation algorithms},
  author={Sarwar, Badrul and Karypis, George and Konstan, Joseph and Riedl, John},
  booktitle={Proceedings of the 10th international conference on World Wide Web},
  pages={285--295},
  year={2001}
}

@article{koren2009matrix,
  title={Matrix factorization techniques for recommender systems},
  author={Koren, Yehuda and Bell, Robert and Volinsky, Chris},
  journal={Computer},
  volume={42},
  number={8},
  pages={30--37},
  year={2009},
  publisher={IEEE}
}

@inproceedings{hu2008collaborative,
  title={Collaborative filtering for implicit feedback datasets},
  author={Hu, Yifan and Koren, Yehuda and Volinsky, Chris},
  booktitle={2008 Eighth IEEE international conference on data mining},
  pages={263--272},
  year={2008},
  organization={Ieee}
}

@article{kula2015metadata,
  title={Metadata embeddings for user and item cold-start recommendations},
  author={Kula, Maciej},
  journal={arXiv preprint arXiv:1507.08439},
  year={2015}
}

@inproceedings{reimers2019sentence,
  title={Sentence-bert: Sentence embeddings using siamese bert-networks},
  author={Reimers, Nils and Gurevych, Iryna},
  booktitle={Proceedings of the 2019 conference on empirical methods in natural language processing and the 9th international joint conference on natural language processing (EMNLP-IJCNLP)},
  pages={3982--3992},
  year={2019}
}

@inproceedings{park2023generative,
  title={Generative agents: Interactive simulacra of human behavior},
  author={Park, Joon Sung and O'Brien, Joseph and Cai, Carrie Jun and Morris, Meredith Ringel and Liang, Percy and Bernstein, Michael S},
  booktitle={Proceedings of the 36th annual acm symposium on user interface software and technology},
  pages={1--22},
  year={2023}
}

@article{jarvelin2002cumulated,
  title={Cumulated gain-based evaluation of IR techniques},
  author={J{\"a}rvelin, Kalervo and Kek{\"a}l{\"a}inen, Jaana},
  journal={ACM Transactions on Information Systems (TOIS)},
  volume={20},
  number={4},
  pages={422--446},
  year={2002},
  publisher={ACM New York, NY, USA}
}

@inproceedings{schein2002methods,
  title={Methods and metrics for cold-start recommendations},
  author={Schein, Andrew I and Popescul, Alexandrin and Ungar, Lyle H and Pennock, David M},
  booktitle={Proceedings of the 25th annual international ACM SIGIR conference on Research and development in information retrieval},
  pages={253--260},
  year={2002}
}

@article{campos2014time,
  title={Time-aware recommender systems: a comprehensive survey and analysis of existing evaluation protocols},
  author={Campos, Pedro G and D{\'\i}ez, Fernando and Cantador, Iv{\'a}n},
  journal={User Modeling and User-Adapted Interaction},
  volume={24},
  number={1},
  pages={67--119},
  year={2014},
  publisher={Springer}
}

@inproceedings{he2017neural,
  title={Neural collaborative filtering},
  author={He, Xiangnan and Liao, Lizi and Zhang, Hanwang and Nie, Liqiang and Hu, Xia and Chua, Tat-Seng},
  booktitle={Proceedings of the 26th international conference on world wide web},
  pages={173--182},
  year={2017}
}

@inproceedings{liang2018variational,
  title={Variational autoencoders for collaborative filtering},
  author={Liang, Dawen and Krishnan, Rahul G and Hoffman, Matthew D and Jebara, Tony},
  booktitle={Proceedings of the 2018 world wide web conference},
  pages={689--698},
  year={2018}
}

@inproceedings{rendle2010factorization,
  title={Factorization machines},
  author={Rendle, Steffen},
  booktitle={2010 IEEE International conference on data mining},
  pages={995--1000},
  year={2010},
  organization={IEEE}
}

@inproceedings{cremonesi2010performance,
  title={Performance of recommender algorithms on top-n recommendation tasks},
  author={Cremonesi, Paolo and Koren, Yehuda and Turrin, Roberto},
  booktitle={Proceedings of the fourth ACM conference on Recommender systems},
  pages={39--46},
  year={2010}
}

@article{deshpande2004item,
  title={Item-based top-n recommendation algorithms},
  author={Deshpande, Mukund and Karypis, George},
  journal={ACM Transactions on Information Systems (TOIS)},
  volume={22},
  number={1},
  pages={143--177},
  year={2004},
  publisher={ACM New York, NY, USA}
}

@article{bobadilla2013recommender,
  title={Recommender systems survey},
  author={Bobadilla, Jes{\'u}s and Ortega, Fernando and Hernando, Antonio and Guti{\'e}rrez, Abraham},
  journal={Knowledge-based systems},
  volume={46},
  pages={109--132},
  year={2013},
  publisher={Elsevier}
}

@article{zhang2019deep,
  title={Deep learning based recommender system: A survey and new perspectives},
  author={Zhang, Shuai and Yao, Lina and Sun, Aixin and Tay, Yi},
  journal={ACM computing surveys (CSUR)},
  volume={52},
  number={1},
  pages={1--38},
  year={2019},
  publisher={ACM New York, NY, USA}
}

@article{herlocker2004evaluating,
  title={Evaluating collaborative filtering recommender systems},
  author={Herlocker, Jonathan L and Konstan, Joseph A and Terveen, Loren G and Riedl, John T},
  journal={ACM Transactions on Information Systems (TOIS)},
  volume={22},
  number={1},
  pages={5--53},
  year={2004},
  publisher={ACM New York, NY, USA}
}

@article{wang2024survey,
  title={A survey on large language model based autonomous agents},
  author={Wang, Lei and Ma, Chen and Feng, Xueyang and Zhang, Zeyu and Yang, Hao and Zhang, Jingsen and Chen, Zhiyuan and Tang, Jiakai and Chen, Xu and Lin, Yankai and others},
  journal={Frontiers of Computer Science},
  volume={18},
  number={6},
  pages={186345},
  year={2024},
  publisher={Springer}
}

@article{durante2024agent,
  title={Agent ai: Surveying the horizons of multimodal interaction},
  author={Durante, Zane and Huang, Qiuyuan and Wake, Naoki and Gong, Ran and Park, Jae Sung and Sarkar, Bidipta and Taori, Rohan and Noda, Yusuke and Terzopoulos, Demetri and Choi, Yejin and others},
  journal={arXiv preprint arXiv:2401.03568},
  year={2024}
}

@article{gao2023chat,
  title={Chat-rec: Towards interactive and explainable llms-augmented recommender system},
  author={Gao, Yunfan and Sheng, Tao and Xiang, Youlin and Xiong, Yun and Wang, Haofen and Zhang, Jiawei},
  journal={arXiv preprint arXiv:2303.14524},
  year={2023}
}

@inproceedings{bao2023tallrec,
  title={Tallrec: An effective and efficient tuning framework to align large language model with recommendation},
  author={Bao, Keqin and Zhang, Jizhi and Zhang, Yang and Wang, Wenjie and Feng, Fuli and He, Xiangnan},
  booktitle={Proceedings of the 17th ACM conference on recommender systems},
  pages={1007--1014},
  year={2023}
}

@article{zhao2024recommender,
  title={Recommender systems in the era of large language models (llms)},
  author={Zhao, Zihuai and Fan, Wenqi and Li, Jiatong and Liu, Yunqing and Mei, Xiaowei and Wang, Yiqi and Wen, Zhen and Wang, Fei and Zhao, Xiangyu and Tang, Jiliang and others},
  journal={IEEE Transactions on Knowledge and Data Engineering},
  volume={36},
  number={11},
  pages={6889--6907},
  year={2024},
  publisher={IEEE}
}

@article{su2009survey,
  title={A survey of collaborative filtering techniques},
  author={Su, Xiaoyuan and Khoshgoftaar, Taghi M},
  journal={Advances in artificial intelligence},
  volume={2009},
  number={1},
  pages={421425},
  year={2009},
  publisher={Wiley Online Library}
}

@article{moltbook2026,
  author  = {Gautam, Sushant and Olstad, Annika W and Pettersen, Klas H and Riegler, Michael A},
  title   = {The Moltbook Observatory Archive: an incremental dataset of agent-only social network activity},
  journal = {arXiv preprint arXiv:2605.13860},
  year    = {2026},
}

@article{linden2003amazon,
  title={Amazon. com recommendations: Item-to-item collaborative filtering},
  author={Linden, Greg and Smith, Brent and York, Jeremy},
  journal={IEEE Internet computing},
  volume={7},
  number={1},
  pages={76--80},
  year={2003},
  publisher={IEEE}
}

@incollection{ricci2010introduction,
  title={Introduction to recommender systems handbook},
  author={Ricci, Francesco and Rokach, Lior and Shapira, Bracha},
  booktitle={Recommender systems handbook},
  pages={1--35},
  year={2010},
  publisher={Springer}
}

@article{gunawardana2009survey,
  title={A survey of accuracy evaluation metrics of recommendation tasks.},
  author={Gunawardana, Asela and Shani, Guy},
  journal={Journal of Machine Learning Research},
  volume={10},
  number={12},
  year={2009}
}

@article{burke2002hybrid,
  title={Hybrid recommender systems: Survey and experiments},
  author={Burke, Robin},
  journal={User modeling and user-adapted interaction},
  volume={12},
  number={4},
  pages={331--370},
  year={2002},
  publisher={Springer}
}

@inproceedings{xi2024towards,
  title={Towards open-world recommendation with knowledge augmentation from large language models},
  author={Xi, Yunjia and Liu, Weiwen and Lin, Jianghao and Cai, Xiaoling and Zhu, Hong and Zhu, Jieming and Chen, Bo and Tang, Ruiming and Zhang, Weinan and Yu, Yong},
  booktitle={Proceedings of the 18th ACM Conference on Recommender Systems},
  pages={12--22},
  year={2024}
}

@article{zimdars2013using,
  title={Using temporal data for making recommendations},
  author={Zimdars, Andrew and Chickering, David Maxwell and Meek, Christopher},
  journal={arXiv preprint arXiv:1301.2320},
  year={2013}
}

@inproceedings{kang2018self,
  title={Self-attentive sequential recommendation},
  author={Kang, Wang-Cheng and McAuley, Julian},
  booktitle={2018 IEEE international conference on data mining (ICDM)},
  pages={197--206},
  year={2018},
  organization={IEEE}
}

@article{goldberg1992using,
  title={Using collaborative filtering to weave an information tapestry},
  author={Goldberg, David and Nichols, David and Oki, Brian M and Terry, Douglas},
  journal={Communications of the ACM},
  volume={35},
  number={12},
  pages={61--70},
  year={1992},
  publisher={ACM New York, NY, USA}
}

@article{harper2015movielens,
  title={The movielens datasets: History and context},
  author={Harper, F Maxwell and Konstan, Joseph A},
  journal={Acm transactions on interactive intelligent systems (tiis)},
  volume={5},
  number={4},
  pages={1--19},
  year={2015},
  publisher={Acm New York, NY, USA}
}

@inproceedings{mcauley2015image,
  title={Image-based recommendations on styles and substitutes},
  author={McAuley, Julian and Targett, Christopher and Shi, Qinfeng and Van Den Hengel, Anton},
  booktitle={Proceedings of the 38th international ACM SIGIR conference on research and development in information retrieval},
  pages={43--52},
  year={2015}
}

@inproceedings{he2020lightgcn,
  title={Lightgcn: Simplifying and powering graph convolution network for recommendation},
  author={He, Xiangnan and Deng, Kuan and Wang, Xiang and Li, Yan and Zhang, Yongdong and Wang, Meng},
  booktitle={Proceedings of the 43rd International ACM SIGIR conference on research and development in Information Retrieval},
  pages={639--648},
  year={2020}
}

@inproceedings{he2014practical,
  title={Practical lessons from predicting clicks on ads at facebook},
  author={He, Xinran and Pan, Junfeng and Jin, Ou and Xu, Tianbing and Liu, Bo and Xu, Tao and Shi, Yanxin and Atallah, Antoine and Herbrich, Ralf and Bowers, Stuart and others},
  booktitle={Proceedings of the eighth international workshop on data mining for online advertising},
  pages={1--9},
  year={2014}
}

@inproceedings{sun2019bert4rec,
  title={BERT4Rec: Sequential recommendation with bidirectional encoder representations from transformer},
  author={Sun, Fei and Liu, Jun and Wu, Jian and Pei, Changhua and Lin, Xiao and Ou, Wenwu and Jiang, Peng},
  booktitle={Proceedings of the 28th ACM international conference on information and knowledge management},
  pages={1441--1450},
  year={2019}
}

@inproceedings{xia2023transact,
  title={Transact: Transformer-based realtime user action model for recommendation at pinterest},
  author={Xia, Xue and Eksombatchai, Pong and Pancha, Nikil and Badani, Dhruvil Deven and Wang, Po-Wei and Gu, Neng and Joshi, Saurabh Vishwas and Farahpour, Nazanin and Zhang, Zhiyuan and Zhai, Andrew},
  booktitle={Proceedings of the 29th ACM SIGKDD Conference on Knowledge Discovery and Data Mining},
  pages={5249--5259},
  year={2023}
}

@inproceedings{zhou2018deep,
  title={Deep interest network for click-through rate prediction},
  author={Zhou, Guorui and Zhu, Xiaoqiang and Song, Chenru and Fan, Ying and Zhu, Han and Ma, Xiao and Yan, Yanghui and Jin, Junqi and Li, Han and Gai, Kun},
  booktitle={Proceedings of the 24th ACM SIGKDD international conference on knowledge discovery \& data mining},
  pages={1059--1068},
  year={2018}
}

@article{deng2025onerec,
  title={Onerec: Unifying retrieve and rank with generative recommender and iterative preference alignment},
  author={Deng, Jiaxin and Wang, Shiyao and Cai, Kuo and Ren, Lejian and Hu, Qigen and Ding, Weifeng and Luo, Qiang and Zhou, Guorui},
  journal={arXiv preprint arXiv:2502.18965},
  year={2025}
}

@article{huang2025towards,
  title={Towards agentic recommender systems in the era of multimodal large language models},
  author={Huang, Chengkai and Wu, Junda and Xia, Yu and Yu, Zixu and Wang, Ruhan and Yu, Tong and Zhang, Ruiyi and Rossi, Ryan A and Kveton, Branislav and Zhou, Dongruo and others},
  journal={arXiv preprint arXiv:2503.16734},
  year={2025}
}

@article{zhai2024actions,
  title={Actions speak louder than words: Trillion-parameter sequential transducers for generative recommendations},
  author={Zhai, Jiaqi and Liao, Lucy and Liu, Xing and Wang, Yueming and Li, Rui and Cao, Xuan and Gao, Leon and Gong, Zhaojie and Gu, Fangda and He, Michael and others},
  journal={arXiv preprint arXiv:2402.17152},
  year={2024}
}

@article{liang2026generative,
  title={Generative Reasoning Re-ranker},
  author={Liang, Mingfu and Li, Yufei and Xu, Jay and Asadi, Kavosh and Liu, Xi and Gu, Shuo and Rangadurai, Kaushik and Shyu, Frank and Wang, Shuaiwen and Yang, Song and others},
  journal={arXiv preprint arXiv:2602.07774},
  year={2026}
}

@article{ashery2025emergent,
  title={Emergent social conventions and collective bias in LLM populations},
  author={Ashery, Ariel Flint and Aiello, Luca Maria and Baronchelli, Andrea},
  journal={Science Advances},
  volume={11},
  number={20},
  pages={eadu9368},
  year={2025},
  publisher={American Association for the Advancement of Science}
}

@article{chen2026ai,
  title={AI agent behavioral science},
  author={Chen, Lin and Zhang, Yunke and Feng, Jie and Chai, Haoye and Zhang, Honglin and Fan, Bingbing and Ma, Yibo and Zhang, Shiyuan and Li, Nian and Liu, Tianhui and others},
  journal={Humanities and Social Sciences Communications},
  year={2026},
  publisher={Palgrave}
}

@article{jiang2026humans,
  title={" Humans welcome to observe": A First Look at the Agent Social Network Moltbook},
  author={Jiang, Yukun and Zhang, Yage and Shen, Xinyue and Backes, Michael and Zhang, Yang},
  journal={arXiv preprint arXiv:2602.10127},
  year={2026}
}

@article{zhang2026agents,
  title={Agents in the wild: Safety, society, and the illusion of sociality on moltbook},
  author={Zhang, Yunbei and Mei, Kai and Liu, Ming and Wang, Janet and Metaxas, Dimitris N and Wang, Xiao and Hamm, Jihun and Ge, Yingqiang},
  journal={arXiv preprint arXiv:2602.13284},
  year={2026}
}

@article{price2026let,
  title={Let there be claws: An early social network analysis of ai agents on moltbook},
  author={Price, Henry CW and AlMuhanna, H and Bassani, PM and Ho, M and Evans, TS},
  journal={arXiv preprint arXiv:2602.20044},
  year={2026}
}

@article{williams2026form,
  title={Form or function? early dynamics of the moltbook ai social media network},
  author={Williams, Nigel and Ferdinand, Nicole},
  journal={ROBONOMICS: The Journal of the Automated Economy},
  volume={7},
  pages={90--90},
  year={2026}
}

@inproceedings{mcauley2013hidden,
  title={Hidden factors and hidden topics: understanding rating dimensions with review text},
  author={McAuley, Julian and Leskovec, Jure},
  booktitle={Proceedings of the 7th ACM conference on Recommender systems},
  pages={165--172},
  year={2013}
}

@article{shinn2023reflexion,
  title={Reflexion: Language agents with verbal reinforcement learning},
  author={Shinn, Noah and Cassano, Federico and Gopinath, Ashwin and Narasimhan, Karthik and Yao, Shunyu},
  journal={Advances in neural information processing systems},
  volume={36},
  pages={8634--8652},
  year={2023}
}

@article{cresci2020decade,
  title={A decade of social bot detection},
  author={Cresci, Stefano},
  journal={Communications of the ACM},
  volume={63},
  number={10},
  pages={72--83},
  year={2020},
  publisher={ACM New York, NY, USA}
}

@article{yao2022react,
  title={React: Synergizing reasoning and acting in language models},
  author={Yao, Shunyu and Zhao, Jeffrey and Yu, Dian and Du, Nan and Shafran, Izhak and Narasimhan, Karthik and Cao, Yuan},
  journal={arXiv preprint arXiv:2210.03629},
  year={2022}
}

@inproceedings{yang2020scalable,
  title={Scalable and generalizable social bot detection through data selection},
  author={Yang, Kai-Cheng and Varol, Onur and Hui, Pik-Mai and Menczer, Filippo},
  booktitle={Proceedings of the AAAI conference on artificial intelligence},
  volume={34},
  number={01},
  pages={1096--1103},
  year={2020}
}

@article{xi2025rise,
  title={The rise and potential of large language model based agents: A survey},
  author={Xi, Zhiheng and Chen, Wenxiang and Guo, Xin and He, Wei and Ding, Yiwen and Hong, Boyang and Zhang, Ming and Wang, Junzhe and Jin, Senjie and Zhou, Enyu and others},
  journal={Science China Information Sciences},
  volume={68},
  number={2},
  pages={121101},
  year={2025},
  publisher={Springer}
}

@article{bogina2023considering,
  title={Considering temporal aspects in recommender systems: a survey: V. Bogina et al.},
  author={Bogina, Veronika and Kuflik, Tsvi and Jannach, Dietmar and Bielikova, Maria and Kompan, Michal and Trattner, Christoph},
  journal={User Modeling and User-Adapted Interaction},
  volume={33},
  number={1},
  pages={81--119},
  year={2023},
  publisher={Springer}
}

@article{vinagre2015overview,
  title={An overview on the exploitation of time in collaborative filtering},
  author={Vinagre, Jo{\~a}o and Jorge, Al{\'\i}pio M{\'a}rio and Gama, Jo{\~a}o},
  journal={Wiley interdisciplinary reviews: Data mining and knowledge discovery},
  volume={5},
  number={5},
  pages={195--215},
  year={2015},
  publisher={Wiley Online Library}
}

@article{li2024recent,
  title={Recent developments in recommender systems: A survey},
  author={Li, Yang and Liu, Kangbo and Satapathy, Ranjan and Wang, Suhang and Cambria, Erik},
  journal={IEEE Computational Intelligence Magazine},
  volume={19},
  number={2},
  pages={78--95},
  year={2024},
  publisher={IEEE}
}

@incollection{lombardi2022ai,
  title={AI-enabled bot and social media: A survey of tools, techniques, and platforms for the arms race},
  author={Lombardi, Flavio and Caprolu, Maurantonio and Di Pietro, Roberto},
  booktitle={Mixed methods perspectives on communication and social media research},
  pages={255--269},
  year={2022},
  publisher={Routledge}
}

@article{shin2026ai,
  title={AI-Gram: When Visual Agents Interact in a Social Network},
  author={Shin, Andrew},
  journal={arXiv preprint arXiv:2604.21446},
  year={2026}
}

@article{abou2025agentic,
  title={Agentic AI: a comprehensive survey of architectures, applications, and future directions},
  author={Abou Ali, Mohamad and Dornaika, Fadi and Charafeddine, Jinan},
  journal={Artificial Intelligence Review},
  volume={59},
  number={1},
  pages={11},
  year={2025},
  publisher={Springer}
}

@article{liu2014new,
  title={A new user similarity model to improve the accuracy of collaborative filtering},
  author={Liu, Haifeng and Hu, Zheng and Mian, Ahmad and Tian, Hui and Zhu, Xuzhen},
  journal={Knowledge-based systems},
  volume={56},
  pages={156--166},
  year={2014},
  publisher={Elsevier}
}

@article{tang2013social,
  title={Social recommendation: a review},
  author={Tang, Jiliang and Hu, Xia and Liu, Huan},
  journal={Social network analysis and mining},
  volume={3},
  number={4},
  pages={1113--1133},
  year={2013},
  publisher={Springer}
}

@article{rafailidis2015modeling,
  title={Modeling users preference dynamics and side information in recommender systems},
  author={Rafailidis, Dimitrios and Nanopoulos, Alexandros},
  journal={IEEE Transactions on Systems, Man, and Cybernetics: Systems},
  volume={46},
  number={6},
  pages={782--792},
  year={2015},
  publisher={IEEE}
}

@inproceedings{carroll2021estimating,
  title={Estimating and penalizing preference shift in recommender systems},
  author={Carroll, Micah and Hadfield-Menell, Dylan and Russell, Stuart and Dragan, Anca},
  booktitle={Proceedings of the 15th ACM Conference on Recommender Systems},
  pages={661--667},
  year={2021}
}

@inproceedings{wang2019neural,
  title={Neural graph collaborative filtering},
  author={Wang, Xiang and He, Xiangnan and Wang, Meng and Feng, Fuli and Chua, Tat-Seng},
  booktitle={Proceedings of the 42nd international ACM SIGIR conference on Research and development in Information Retrieval},
  pages={165--174},
  year={2019}
}

@article{lirise,
  title={The Rise of Autonomous AI Agents: A Comprehensive Survey of OpenClaw—Architecture, Security, Ecosystem, and Beyond},
  author={Li, Siyuan and Shu, Peng and Yu, Churan and Wang, Peilong and Zhang, Ruidong and Guo, Bowen and Li, Xinliang and Yan, Ruiyu and Zidan, Arif Hassan and Pan, Yi and others}
}

@article{zhou2026externalization,
  title={Externalization in llm agents: A unified review of memory, skills, protocols and harness engineering},
  author={Zhou, Chenyu and Chai, Huacan and Chen, Wenteng and Guo, Zihan and Shan, Rong and Song, Yuanyi and Xu, Tianyi and Yang, Yingxuan and Yu, Aofan and Zhang, Weiming and others},
  journal={arXiv preprint arXiv:2604.08224},
  year={2026}
}

@inproceedings{sakai2006evaluating,
  title={Evaluating evaluation metrics based on the bootstrap},
  author={Sakai, Tetsuya},
  booktitle={Proceedings of the 29th Annual International ACM SIGIR Conference on Research and Development in Information Retrieval (SIGIR)},
  pages={525--532},
  year={2006}
}

@inproceedings{smucker2007comparison,
  title={A comparison of statistical significance tests for information retrieval evaluation},
  author={Smucker, Mark D. and Allan, James and Carterette, Ben},
  booktitle={Proceedings of the 16th ACM Conference on Information and Knowledge Management (CIKM)},
  pages={623--632},
  year={2007}
}

@incollection{shani2010evaluating,
  title={Evaluating recommendation systems},
  author={Shani, Guy and Gunawardana, Asela},
  booktitle={Recommender systems handbook},
  pages={257--297},
  year={2010},
  publisher={Springer}
}

@article{shen2026agentic,
  title={Agentic Recommender System with Hierarchical Belief-State Memory},
  author={Shen, Xiang and Zhou, Yuhang and Wu, Yifan and Zhao, Zhuokai and Lin, Siyu and Huang, Lei and Zhong, Qianqian and Zhang, Lizhu and Zhang, Benyu and Fan, Xiangjun and others},
  journal={arXiv preprint arXiv:2605.14401},
  year={2026}
}

@article{nguyen2026amem4rec,
  title={Amem4rec: Leveraging cross-user similarity for memory evolution in agentic llm recommenders},
  author={Nguyen, Minh-Duc and Kieu, Hai-Dang and Le, Dung D},
  journal={arXiv preprint arXiv:2602.08837},
  year={2026}
}

@inproceedings{liu2026recoworld,
  title={Recoworld: Building simulated environments for agentic recommender systems},
  author={Liu, Fei and Lin, Xinyu and Yu, Hanchao and Wu, Mingyuan and Wang, Jianyu and Zhang, Qiang and Zhao, Zhuokai and Xia, Yinglong and Zhang, Yao and Li, Weiwei and others},
  booktitle={Companion Proceedings of the ACM Web Conference 2026},
  pages={650--659},
  year={2026}
}

@article{mou2026individual,
  title={From individual to society: A survey on social simulation driven by large language model-based agents},
  author={Mou, Xinyi and Ding, Xuanwen and He, Qi and Wang, Liang and Liang, Jingcong and Zhang, Xinnong and Sun, Libo and Lin, Jiayu and Zhou, Jie and Xuanjing, Huang and others},
  journal={ACM Computing Surveys},
  volume={58},
  number={11},
  pages={1--41},
  year={2026},
  publisher={ACM New York, NY}
}

@inproceedings{sun2026h,
  title={H-mem: Hierarchical memory for high-efficiency long-term reasoning in llm agents},
  author={Sun, Haoran and Zeng, Shaoning and Zhang, Bob},
  booktitle={Proceedings of the 19th Conference of the European Chapter of the Association for Computational Linguistics (Volume 1: Long Papers)},
  pages={341--350},
  year={2026}
}

@article{xu2026mem,
  title={A-mem: Agentic memory for llm agents},
  author={Xu, Wujiang and Liang, Zujie and Mei, Kai and Gao, Hang and Tan, Juntao and Zhang, Yongfeng},
  journal={Advances in Neural Information Processing Systems},
  volume={38},
  pages={17577--17604},
  year={2026}
}

@inproceedings{chen2026memrec,
  title={Memrec: Collaborative memory-augmented agentic recommender system},
  author={Chen, Weixin and Zhao, Yuhan and Huang, Jingyuan and Ye, Zihe and Ju, Mingxuan and Zhao, Tong and Shah, Neil and Chen, Li and Zhang, Yongfeng},
  booktitle={Proceedings of the 64th Annual Meeting of the Association for Computational Linguistics (Volume 1: Long Papers)},
  pages={44515--44544},
  year={2026}
}

@inproceedings{resnick1994grouplens,
  title={Grouplens: An open architecture for collaborative filtering of netnews},
  author={Resnick, Paul and Iacovou, Neophytos and Suchak, Mitesh and Bergstrom, Peter and Riedl, John},
  booktitle={Proceedings of the 1994 ACM conference on Computer supported cooperative work},
  pages={175--186},
  year={1994}
}

\end{document}